\documentclass[preprint,3p]{elsarticle}
\usepackage{subfig}
\usepackage{graphicx}
\usepackage{amsmath}

\makeatletter
\def\ps@pprintTitle{
  \let\@oddhead\@empty
  \let\@evenhead\@empty
  \def\@oddfoot{\footnotesize\itshape\hfill\today}
  \let\@evenfoot\@oddfoot
}
\makeatother

\begin{document}

\title{Simulation of incompressible two-phase flow in porous media with large timesteps}
\author[]{Daniel A. Cogswell\corref{cor1}}
\ead{cogswell@alum.mit.edu}
\cortext[cor1]{Corresponding author}
\author[]{Michael L. Szulczewski}
\address{Aramco Services Company: Aramco Research Center-Boston, 400 Technology Square, Cambridge, Massachusetts 02139, USA}

\begin{abstract}
Multiphase flow in porous media occurs in several disciplines including petroleum reservoir engineering, petroleum systems' analysis, and CO$_2$ sequestration.  While simulations often use a fully implicit discretization to increase the time step size, restrictions on the time step often exist due to non-convergence of the nonlinear solver (e.g. Newton's method).  Here this problem is addressed for the Buckley-Leverett equations, which model incompressible, immiscible, two-phase flow with no capillary potential. The equations are recast as a gradient flow using the phase-field method, and a convex energy splitting scheme is applied to enable large timesteps, even for high degrees of heterogeneity in permeability and viscosity.  By using the phase-field formulation as a homotopy map, the underlying hyperbolic flow equations can be solved with large timesteps. For a heterogeneous test problem, the new homotopy method allows the timestep to be increased by more than six orders of magnitude relative to the unmodified equations while maintaining convergence.
\end{abstract}

\begin{keyword}
Reservoir simulation, two-phase flow, phase-field method, homotopy, multigrid
\end{keyword}

\maketitle

\section{Introduction}
Timestep restrictions during simulation of fluid flow through extremely heterogeneous porous media remain a significant limiting factor in petroleum reservoir models \cite{Fung2008}. The equations of flow in porous media \cite{Buckley1942,Peaceman1977,Pinder2008} suffer from severe timestep restrictions as the permeability and viscosity contrast become increasingly heterogeneous \cite{Jenny2009,Younis2010,Li2015}, even when solved with a fully implicit discretization. Often there is no choice but to significantly reduce the timestep in order to regain convergence.  Improved numerical methods have helped \cite{Dawson1997,Jenny2009,Younis2010,Li2015}, but the problem persists because it originates from the shape of the fractional flow function which leads to divergence of the iterative method used to solve the discrete nonlinear equations.

Either the numerical method or the equations themselves must be modified to achieve convergence for large timesteps. Previous efforts have focused on the numerical method with the use of line-search \cite{Dawson1997}, trust-region \cite{Jenny2009,Li2015}, or continuation methods \cite{Younis2010}. Here we focus on regularizing the equations themselves with the addition of an energy constraint. Gradient flows, where evolution equations are derived from energy functionals, offer an attractive alternative formulation for the flow equations. In particular, the phase-field method has emerged as an effective way to solve free boundary problems without explicitly tracking interfaces, and models of fluid flow have been rigorously derived from thermodynamic principles \cite{Cahn1958,Lowengrub1998,Anderson1998,Lee2002}. Phase-field methods offer numerical advantages as well, since they guarantee monotonically decreasing energy of the solution. For example, Feng and Wise \cite{Wise2010,Feng2012} recently showed that a Cahn-Hilliard-Darcy system has an unconditionally energy-stable and unconditionally uniquely solvable discretization. The combination of these two properties implies that equations solved using Newton's method will always converge.

The phase-field method has had limited application to flow in porous media, however, and convex energy splitting has not yet been applied to address timestep restrictions. Cueto-Felgueroso and Juanes developed a phase-field model of unsaturated flow, and successfully simulated gravity fingering in soil \cite{Cueto-Felgueroso2008,Cueto-Felgueroso2009} and  bubble motion in a capillary tube \cite{Cueto-Felgueroso2012,Cueto-Felgueroso2014}. Chen and Yan \cite{Chen2015} applied a phase-field model of fluid displacement to viscous fingering in a heterogeneous permeability field, but did not employ fractional flow functions for unsaturated flow.

In this paper, the phase-field method is extended to multiphase flow in porous media.  For simplicity, two-phase incompressible flow of immiscible fluids is considered with no capillary potential, relative permeability hysteresis, or gravitational effects.  The equations that model this flow are often referred to as the Buckley-Leverett equations~\cite{Buckley1942,Pinder2008}.  Using the phase-field method, this system is augmented to include a macroscopic surface tension that is analogous to an artificial viscosity~\cite{VonNeumann1950}, and a semi-implicit temporal discretization is employed based on a convex splitting of the free energy.  In a related system---two-phase flow in a Hele-Shaw cell---this discretization leads to a system of equations that has been proven to be unconditionally and uniquely solvable~\cite{Wise2010,Feng2012}. Neither unconditional nor unique solvability are proven here for the modified Buckley-Leverett system. Instead, arbitrarily large time steps are demonstrated on a test problem with permeability heterogeneity and different fluid viscosities. Homotopy \cite{Chow1978,Watson1989} is used to eliminate the physical effects of macroscopic surface tension on the solution; surface tension is progressively decreased after each iteration of the nonlinear solver until the surface tension terms are negligibly small or zero and a solution to the original equations is obtained. With this approach it is possible to solve the unmodified set of equations using timesteps several orders of magnitude larger.
 
\section{Equations}
The traditional equations for incompressible, immiscible two-phase flow in porous media with no capillary potential\footnote{Here capillary potential refers to a pressure difference between phases that causes flow, which is separate from the Laplace pressure due to the curvature of an interface between fluids.} or gravitational effects are \cite{Peaceman1977}:
\begin{subequations}
\begin{equation}
 \phi\frac{\partial S_i}{\partial t}+\nabla\cdot\left(\mathcal F_i\vec{v}_t\right)=q_i
\end{equation}
\begin{equation}
 \nabla\cdot\vec v_t=q_t
\end{equation}
\begin{equation}
 \vec v_i=-k\lambda_i\vec\nabla p
\end{equation}
 \label{Eq:BL}
\end{subequations}
For a two-phase system of oil and water, $i=o$ denotes the oil phase and $i=w$ the water phase. The first equation is the saturation evolution equation, the second is a continuity equation, and the third is Darcy's Law for flow in a porous medium. The porosity is $\phi$ (assumed here to be constant), $S_i$ is the saturation of phase $i$, $p$ is pressure, $\mathcal F_i$ is the fraction of the flowing stream comprised of phase $i$, $\vec v_t=\vec v_o+\vec v_w$ is the total velocity, $k$ is the permeability tensor of the porous material, and $\lambda_i=\frac{k_{ri}}{\eta_i}$ is the transmissibility of phase $i$ with relative permeability $k_{ri}$ and viscosity $\eta_i$. The source term $q_i=\sum_j q_{ij}\delta(\vec x-\vec x_j)$ is the sum of injection/production rates of phase $i$ over all wells $j$, where $\delta(\vec x-\vec x_j)$ is a Dirac delta function that indicates the location of well $j$. The total injection/production of fluid is $q_t=q_w+q_o$, the sum of water and oil injection/production.

\begin{figure}
 \centering
 \includegraphics[width=.5\columnwidth]{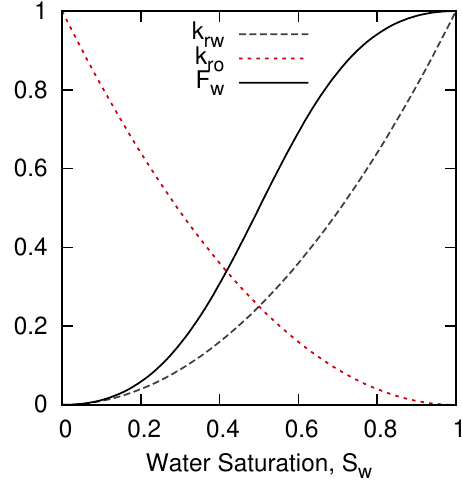}
 \caption{Example relative permeabilities of water and oil, $k_{rw}=S_w^2$ and $k_{ro}=(1-S_w)^2$, and the corresponding fractional flow function for water in a porous material for $\eta_w=\eta_o=1$.}
 \label{Fig:relperm}
\end{figure}

The fractional flow of a phase is the ratio of its transmissibility to the total transmissibility of all phases. A two-phase system is completely characterized by the fractional flow function of one of the phases, which is chosen here to be water:
\begin{equation}
 \mathcal F_w=\frac{\lambda_w}{\lambda_w+\lambda_o}
\end{equation}
The corresponding fractional flow function of oil is simply $\mathcal F_o=1-\mathcal F_w$. An example of such a function and its corresponding relative permeability functions are shown in Fig. \ref{Fig:relperm} in the absence of gravity.  The fractional flow function can be generally characterized as a sigmoidal curve defined between two saturations that place a bound on the range of saturations where flow occurs. In practice the relative permeability functions can often be fit with a power law function \cite{Pinder2008} (pg. 146). If the slope of the fractional flow function approaches zero, fundamental difficulties are encountered with the convergence of Newton-based methods, which will now be discussed.

\subsection{Convergence Analysis}
Newton's method is frequently used to solve discretized versions of the the Buckley-Leverett system. Motivated by work suggesting that the nonlinearity of the discretized equations is dominated by the nonlinearity of the fractional flow function \cite{Jenny2009,Li2015}, the convergence properties of Newton's method applied to the fractional flow function $\mathcal F_w$ illustrated in Fig. \ref{Fig:relperm} will now be analyzed.

The difficulty with Newton's method and fractional flow functions can be appreciated by applying the classical convergence analysis of Kantorovich \cite{Kantorovich1982}, from which it is straightforward to show that Newton's method is not guaranteed to converge from an arbitrary guess for many choices of the fractional flow function. Kantorovich's theorem  (specifically, the version based on the assumptions of Mysovskikh \cite{Kantorovich1982}, Theorem 5, pg. 539) states that for a real function $f(x)$ and an initial guess $x_0$, Newton's method is guaranteed to converge quadratically to a root when $f'(x)$ is invertible and the following condition is obeyed:
\begin{equation}
 \frac{\left|f(x_0)\right|}{\left|f'(x_0)\right|^2}K\le 2
 \label{Eq:Kantorovich}
\end{equation}
where $K=\sup\left|f''(x)\right|$ is the Lipschitz coefficient that places a bound on the second derivative over the domain of interest \cite{Hubbard2009}. Kantorovich's analysis also applies to vector-valued functions \cite{Kantorovich1982} (Theorem 1, pg. 545) if the absolute values in Eq. \ref{Eq:Kantorovich} are replaced with the appropriate vector or matrix norms, $f'(x_0)$ is the Jacobian, and $K$ is the square root of the largest singular value of the Hessian of $f(x_0)$.

\begin{figure}
 \centering
 \includegraphics[width=.55\columnwidth]{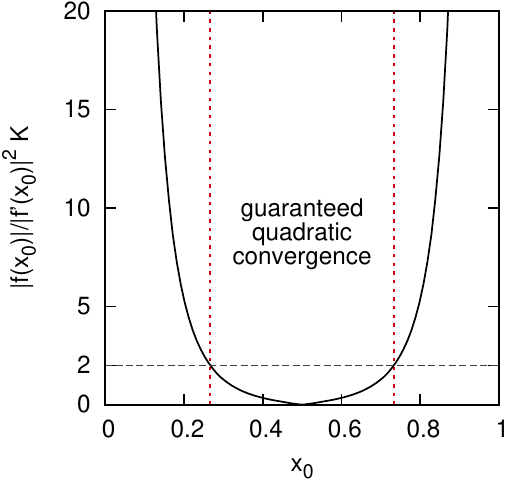}\hspace{.05\columnwidth}
 \caption{The Kantorovich theorem (Eq. \ref{Eq:Kantorovich}) for convergence to a root of $F_w(x)-.5$ from an initial guess $x_0$, where $F_w$ is the fractional flow curve in Fig. 1.}
 \label{Fig:Mysovskikh}
\end{figure}

By examining Eq. \ref{Eq:Kantorovich} it can be reasoned that initial guesses that are either close to the true solution $f(x^*)=0$ or that have a large slope $f'(x_0)$ are likely to converge, while starting points of zero slope where $f'(x_0)=0$ will not converge. Convergence from other starting points will depend on the shape of the function, as is illustrated in Fig. \ref{Fig:Mysovskikh} for finding the root of $\mathcal F_w(x)-.5$ with Newton's method.

Kantorovich's analysis provides motivation for restarting a diverging initial guess from an inflection point where $f''(x_0)=0$, as has been suggested in other work \cite{Jenny2009,Li2015}. Due to the generally sigmoidal shape of fractional flow functions, the inflection point is likely to have a large slope. The condition in Eq. \ref{Eq:Kantorovich} for quadratic convergence is more likely to be met when $f'(x_0)$ is large, but convergence is not guaranteed. However, a convexification of the underlying equations can ensure that that starting points which are far from a solution will not become stuck at points of vanishing slope. This can be accomplished with the addition of an energy functional and the application of variational methods, which ensure that the solution always evolves toward its minimum energy.

\subsection{Macroscopic surface tension}
\begin{figure}
 \centering
 \includegraphics[width=.6\columnwidth]{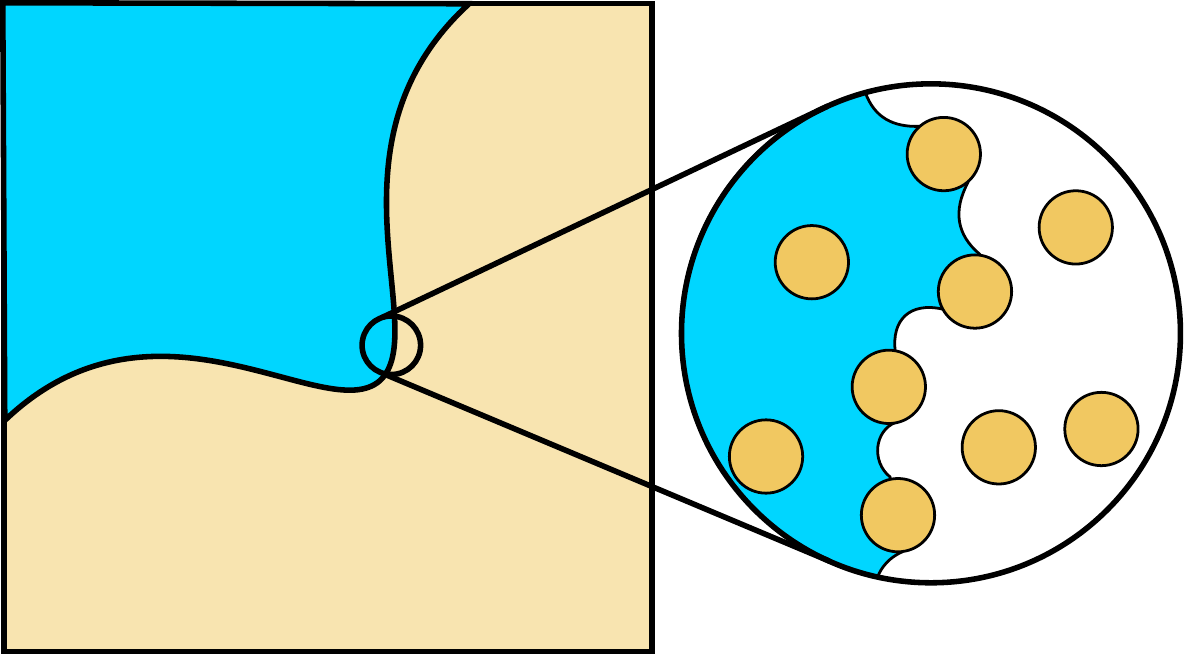}
 \caption{Illustration of the difference between a microscopic interface in a porous material and a macroscopic interface at the Darcy scale.}
 \label{Fig:macroscopic}
\end{figure}
The two-phase flow equations will now be recast as a phase-field model with the introduction of surface tension at the macroscopic scale of Darcy flow. Fig. \ref{Fig:macroscopic} illustrates the difference between the Darcy regime, where the porous material is treated as homogeneous, and the microscopic regime, where surface tension and the heterogeneity of the porous material play microscopic roles. In the Darcy regime, macroscopic surface tension is implemented here with a gradient energy, following the phase-field method. As will be shown, this macroscopic surface tension stabilizes the two-phase flow equations in an analogous way to artificial viscosity, which is often added to fluid dynamics simulations for numerical reasons. Although macroscopic surface tension may have a physical interpretation in some systems, notably gravity fingering in soils where it was incorporated via a gradient energy \cite{Cueto-Felgueroso2008,Cueto-Felgueroso2009}, here macroscopic surface tension is treated as an artificial quantity added to achieve numerical convergence for large timesteps.

Following previous phase-field models of binary fluids \cite{Lowengrub1998,Anderson1998,Lee2002}, a free energy functional for two-phase flow may be defined as:
\begin{equation}
 F=\int \frac 12\rho\left|\vec v\right|^2+Hf(\xi)+\frac 12\kappa |\vec\nabla\xi |^2\, dV
\end{equation}
where $\xi$ is an order parameter that varies between 0 and 1 and indicates which phase is present. The first term is the kinetic energy of the fluid, the second is chemical free energy, and the third is surface energy. $f(\xi)=\xi^2(1-\xi)^2$ is a double-well function, and the gradient energy $\kappa$ and barrier height $H$ are constants related to the surface tension:
\begin{equation}
 \gamma=\sqrt{\frac{\kappa H}{18}}
\end{equation}
and the width of the diffuse interface:
\begin{equation}
 W=\sqrt{\frac{8\kappa}{H}}
\end{equation}
This is by no means the only choice for the double-well function, although it has the advantage of being a polynomial with the simple analytic expressions for $\gamma$ and $\delta$ above \cite{Cahn1958}. The double-well function may need to be altered based on physical considerations.\footnote{For example, if flow does not occur over the entire range of saturations it may be necessary to move the locations of the energy wells with a function of the form $f(S_w)=(S_w-S_{wc})^2(1-S_w-S_{or})^2$, where the position of the wells are specified by $S_{wc}$, the connate water saturation, and $S_{or}$, the residual oil saturation.}

In the Darcy regime, viscous forces dominate over inertial forces and kinetic energy does not contribute to the energy functional \cite{Lee2002}:
\begin{equation}
 F=\int Hf(\xi)+\frac 12\kappa |\vec\nabla\xi |^2\, dV
 \label{Eq:energy_functional}
\end{equation}
Eq. \ref{Eq:energy_functional} may be written in terms of $\gamma$ and $W$ using the relationships in Eq. 5 and 6 as:
\begin{equation}
 F=3\gamma\int \frac{4}{W}f(\xi)+\frac W4|\vec\nabla\xi |^2\, dV
 \label{Eq:energy_functional_gamma}
\end{equation}
which illustrates that the total free energy of the system scales with surface tension. This is an important observation, as $\gamma$ is the parameter that will enable large timesteps with the use of a homotopy method, as discussed in Sec. \ref{Sec:homotopy}.

In porous media the order parameter is identified as the water saturation $\xi=S_w$, the fraction of pore space filled with water, and the oil saturation is defined as $S_o=1-S_w$. The  advection-diffusion equation for incompressible flow becomes:
\begin{subequations}
\begin{equation}
 \phi\frac{\partial S_w}{\partial t}+\nabla\cdot\left(\mathcal F_w\vec{v}_t\right)=\nabla\cdot \left(\frac{k}{\bar\eta}\vec\nabla\hat\mu_w\right)+q_w
 \label{Eq:saturation}
\end{equation}
\begin{equation}
 \nabla\cdot\vec v_t=q_t
\end{equation}
\begin{equation}
 \vec v_i=-k\lambda_i\left[\vec\nabla p+\kappa\nabla\cdot(\vec\nabla S_i\otimes\vec\nabla S_i)\right]
 \label{Eq:Darcy}
\end{equation}
 \label{Eq:CHD}
\end{subequations}
where $\hat\mu_w=\frac{\delta F}{\delta S_w}=Hf'(S_w)-\kappa\nabla^2S_w$ is the variational derivative of Eq. \ref{Eq:energy_functional}, and $\bar\eta=\frac{2\eta_w\eta_o}{\eta_w+\eta_o}$ is the harmonic average of the viscosities of the two fluids. This set of equations models a Cahn-Hilliard-Darcy system (CHD) with nonlinear relative permeability functions incorporated. The new terms that have been added are proportional to the surface tension, and the magnitude of this quantity controls the degree to which the CHD system differs from the porous media equations (Eq.~\ref{Eq:BL}). The new saturation equation (Eq. \ref{Eq:saturation}) is a natural extension of the Buckley-Leverett saturation equation \cite{Buckley1942,Peaceman1977} with a macroscopic surface tension incorporated on the right hand side of the equation according to the phase-field method. The Darcy velocity in Eq. \ref{Eq:CHD} now includes a capillary stress tensor, $\kappa(\vec\nabla S_i\otimes\vec\nabla S_i)$, which is necessary to enforce zero traction at the interface \cite{Lowengrub1998,Anderson1998,Jacqmin1999}. The CHD formulation reduces to the phase-field model of Hele-Shaw cells \cite{Lee2002} for the particular choice of $k_{rw}=S_w$, $k_{ro}=1-S_w$, and $\eta_w=\eta_o$, in which case the gradient energy rigorously depicts the interface between the two fluids.

\section{Numerical methods}
\label{Sec:numerical_methods}
Making an appropriate convex energy splitting (i.e. a semi-implicit discretization) leads to unconditionally energy-stable and unconditionally solvable discretization of gradient flows \cite{Eyre1998,Eyre1998a,Vollmayr-Lee2003}. Each term in the energy functional is assigned to either implicit or explicit treatment based on the sign of the eigenvalues of its Hessian matrix, an approach that has previously been applied to the Cahn-Hilliard-Hele-Shaw system \cite{Wise2010,Feng2012}. Building upon this approach, the splitting technique is applied to the CHD system with nonlinear relative permeability functions, represented by Eq. \ref{Eq:CHD}. The double-well function is split into $f(S_w)=f_c(S_w)+f_e(S_w)$ with a contractive part $f_c''(S_w)>0$ and an expansive part $f_e''(S_w)<0$.

There are many possible ways to choose the convex splitting, and whether there is an optimal choice remains an open question. An obvious first choice of functions is $f_c(S_w)=S_w^2+S_w^4$ and $f_e(S_w)=-2S_w^3$, and this choice was found to work in practice. However, a better choice can be made by considering the Kantorovich convergence criterion, Eq. \ref{Eq:Kantorovich}, which guarantees quadratic convergence. If $f_c(S_w)$ is chosen such that the Hessian of $f'_c(S_w)$ is zero (i.e. $f'''_c(S_w)=0$), the energy function will not contribute to the Lipschitz coefficient in Eq. \ref{Eq:Kantorovich}, and therefore will not adversely affect the convergence rate. The choice of $f_c(S_w)=S_w^2$ and $f_e(S_w)=-2S_w^3+S_w^4$ obeys this criterion and was observed to converge more efficiently, and as a result was the splitting choice used in this work.

The following discretization is then obtained when heterogeneous permeability, fractional flow functions, viscosity contrast, and source terms are included:
\begin{subequations}
\begin{equation}
 \frac{S_w^{n+1}-S_w^n}{\Delta t}+\nabla\cdot\left(\mathcal F_w\vec{v}_t\right)^{n+1}=\nabla\cdot \left(\frac{k}{\bar\eta}\vec\nabla\hat\mu_w^{n+1}\right)+q_w^{n+1}
 \label{Eq:discretized_saturation}
\end{equation}
\begin{equation}
 \hat\mu_w^{n+1}-Hf_c'(S_w^{n+1})+\kappa\nabla^2 S_w^{n+1}=Hf_e'(S_w^n)
 \label{Eq:discretized_mu}
\end{equation}
\begin{equation}
 \nabla\cdot\vec v_t^{n+1}=q_t^{n+1}
 \label{Eq:discretized_continuity}
\end{equation}
 \label{Eq:discretization}
\end{subequations}
The first equation is the discretized saturation equation, the second is an update equation for the diffusion potential $\hat\mu$, and the third is a continuity equation. Since the phase-field term in Eq. \ref{Eq:saturation} has a $\nabla^4S$ term,  it is numerically beneficial to break the equation into two second-order equations, Eq. \ref{Eq:discretized_saturation} and \ref{Eq:discretized_mu}, which can then be treated with standard finite volume methods. Whereas the CHD system (Eq. \ref{Eq:CHD}) has two unknowns, $S_w$ and $p$, the discretized system (Eq. \ref{Eq:discretization}) has three unknowns, $S_w$, $\hat\mu_w$, and $p$.

The Darcy velocity (Eq. \ref{Eq:Darcy})  can be modified for numerical convenience by converting it to the potential form \cite{Jacqmin1999,Feng2006}:
\begin{equation}
 \vec v_i=-k\lambda_i\left[\vec\nabla \bar p-\hat\mu_w\vec\nabla S_w\right]
 \label{Eq:Darcy_potential}
\end{equation}
where $-\hat\mu_w\vec\nabla S_w$ is a continuum forcing term, and the pressure $\bar p$ in Eq. \ref{Eq:Darcy_potential} has been defined as $\bar p=p+Hf(S_w)+\frac 12 \kappa |\vec\nabla S_w|^2$. This form is mathematically equivalent to Eq. \ref{Eq:Darcy} but more numerically suitable since the inversion of an outer product of gradients is avoided.

	Aside from the semi-implicit treatment in Eq. \ref{Eq:discretized_mu} which resulted from the convex energy splitting procedure, all of the remaining terms were treated fully implicitly. Equation ~\ref{Eq:discretization} was discretized in space using the finite volume method~\cite{Leveque2004} and a standard first-order upwind scheme.  Upwind values of the transmissibilities, $\lambda_i$, were chosen based on the sign of $\vec\nabla\bar p - \hat\mu_w\vec\nabla S_w$ at each cell face, and the permeability at each cell face was calculated as the harmonic mean of the permeabilities in adjacent cells. Zero Darcy velocity was enforced at all boundaries by applying Neumann conditions (zero slope) to all variables.

Multigrid methods, previously demonstrated to be an efficient way to solve two-phase flow in porous media with discontinuous permeability \cite{TeiglandFladmark1991,Ersland1993}, were used to solve Eq. \ref{Eq:discretization} (supplemented by Eq. \ref{Eq:Darcy_potential}) for the variables $S$, $\hat\mu$, and $\bar p$. The Full Approximation Scheme (FAS) nonlinear multigrid method was used with a Red-Black Gauss-Seidel smoother \cite{Brandt2011,Trottenberg2001} and  F(2,2) cycles, which were found to be the most efficient. The cell-centered restriction operator (i.e. averaging) was used along with bilinear interpolation \cite{Trottenberg2001}.

\subsection{Homotopy map}
\label{Sec:homotopy}
Although the phase-field formulation offers convergence for large timesteps, it comes at the price of solving a modified physical problem. It is possible, however, to use the phase-field formulation as a homotopy map \cite{Chow1978,Watson1989} so that the Buckely-Leverett problem can be solved directly with large timesteps. Using homotopy, the phase-field solution is continuously transformed into the underlying Buckley-Leverett solution as $\gamma\rightarrow 0$.

By progressively decreasing $\gamma$ during the iterations to a solution at each timestep, it is possible to solve the unmodified Buckely-Leverett equations with large timesteps. After each multigrid cycle, $\gamma$ is multiplied by a constant $w$, where $0<w<1$. Small values of $w$ reach the solution to the unmodified equations in fewer iterations, but a too small $w$ can cause a loss of convergence of the homotopy method. The progression of the guess after each multigrid cycle is illustrated in Fig. \ref{Fig:homotopy} for the 2D test problem discussed in section \ref{Sec:2D_flow}. Although the choice of $\gamma$ and $w$ will likely vary from problem to problem, $w=.5$ was effective for this test problem.

\section{Results}
\subsection{Comparison to the Buckley-Leverett solution}
\begin{figure}[t]
 \subfloat[]{\includegraphics[width=.45\textwidth]{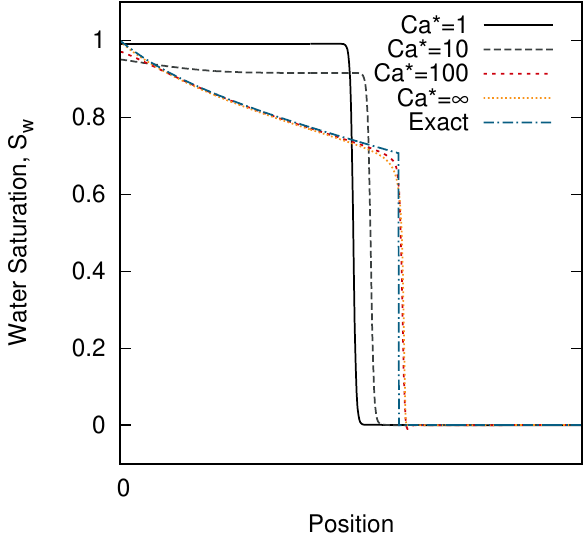}}
 \hspace{.05\textwidth}
 \subfloat[]{\includegraphics[width=.48\textwidth]{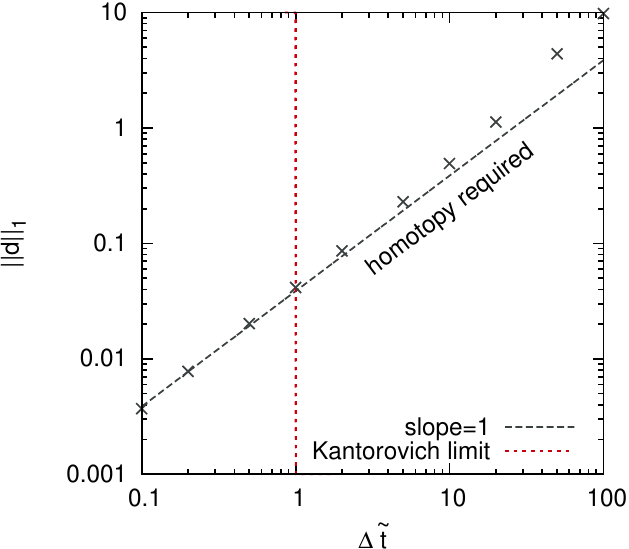}}
 \caption{The effect of macroscopic surface tension and homotopy on the Buckley-Leverett solution in 1D. (a) Comparison of CHD at different capillary numbers with the Buckley-Leverett solution, using the relative permeabilities $k_{rw}=S_w^2$, $k_{ro}=(1-S_w)^2$ and $\eta_w=\eta_o$. All solutions were calculated with $\Delta\tilde t=1$ (i.e. no homotopy), which is the largest stable timestep for $Ca^*=\infty$. (b) Numerical error as a function of timestep size for the Buckley-Leverett solution, computed using the phase-field homotopy method.}
  \label{Fig:Buckley-Leverett}
\end{figure}

A well-known model for the displacement of one fluid by another in porous sand was developed by Buckley and Leverett \cite{Buckley1942}, who analyzed the fractional flow function  (Fig. \ref{Fig:relperm}) and showed that it was possible to derive an analytic solution for the shock that developed. Comparison of the phase-field solution to this analytic solution solution provides one way of quantifying the influence of the macroscopic surface tension $\gamma$ in the CHD model. 

Fig. \ref{Fig:Buckley-Leverett}a compares the Buckley-Leverett solution to the CHD solution for different capillary numbers, with the macroscopic capillary number is defined as:
\begin{equation}
 Ca^*=\frac{\eta\left|\vec v_t\right|}{\gamma}
\end{equation}
In the limit $Ca^*\rightarrow\infty$ (i.e. $\gamma\rightarrow0$) the Buckley-Leverett equations (Eq. \ref{Eq:BL}) are recovered. At large capillary numbers, viscous forces dominate and the CHD solution is very close to the analytic solution. As $Ca^*$ decreases, surface tension begins to play a role. For $Ca^*\le 10$, artificial surface tension strongly impacts the solution, which exhibits a plateau at increasingly larger water saturations with decreasing $Ca^*$. This trend was also observed in the simulations of Cueto-Felgueroso and Juanes \cite{Cueto-Felgueroso2012, Cueto-Felgueroso2014} in the context of bubble motion in a Hele-Shaw cell. For $Ca^*\ge 100$, however, surface tension has a small impact and the solution of the CHD equations agree with the Buckley-Leverett solution.

Although Fig. \ref{Fig:Buckley-Leverett}a shows the Buckley-Leverett solution over a wide range of $Ca^*$ for curiosity, only a small amount of surface tension is actually required to regain numerical convergence with the homotopy progression of Section \ref{Sec:homotopy}. For instance, $\Delta\tilde t$ in Fig. \ref{Fig:Buckley-Leverett} can be increased by several orders of magnitude by starting the homotopy solution with a capillary number of $Ca^*=100$, where surface tension has little effect.

The effect of homotopy on the Buckley-Leverett solution is presented in Fig \ref{Fig:Buckley-Leverett}b, which shows numerical error as a function of timestep size. The error was measured with respect to a discrete reference solution that was calculated using $\Delta\tilde t=.01$ and no homotopy. Figure \ref{Fig:Buckley-Leverett}b demonstrates that the homotopy iterations do not disrupt the expected first-order convergence over several orders of magnitude, even as the timestep is increased beyond the Kantorovich limit. This limit was obtained by solving the vector-valued version of Eq. \ref{Eq:Kantorovich} for the largest value of $\Delta\tilde t$ that does not violate the inequality.

\subsection{Porous flow in 2D}
\label{Sec:2D_flow}

\begin{table}[t]
 \centering
 \begin{tabular}{ccc}
  \hline
  Variable & Description & Value \\
  \hline
  $k_{rw}$ & relative permeability of water &  $S_w^2$ \\
  $k_{ro}$ & relative permeability of oil &  $(1-S_w)^2$ \\
  $\tilde q$ & injection rate of water & $.25$ \\
  $\tilde\eta_w$ & viscosity of water &  1 \\
  $\tilde\eta_o$ & viscosity of oil &  100 \\
  $\tilde\gamma$ & macroscopic surface tension & $2\times10^{-3}$ \\
  $\tilde W$ & interfacial width & 6 \\
  \hline
 \end{tabular}
 \caption{Nondimensional parameters (see \ref{Sec:Nondimensionalization}) for the 2D flow simulation in Fig. \ref{Fig:2D-injection}.}
 \label{Tbl:parameters}
\end{table}

\begin{figure*}[t]
\centering
\subfloat[]{\includegraphics[width=.45\textwidth]{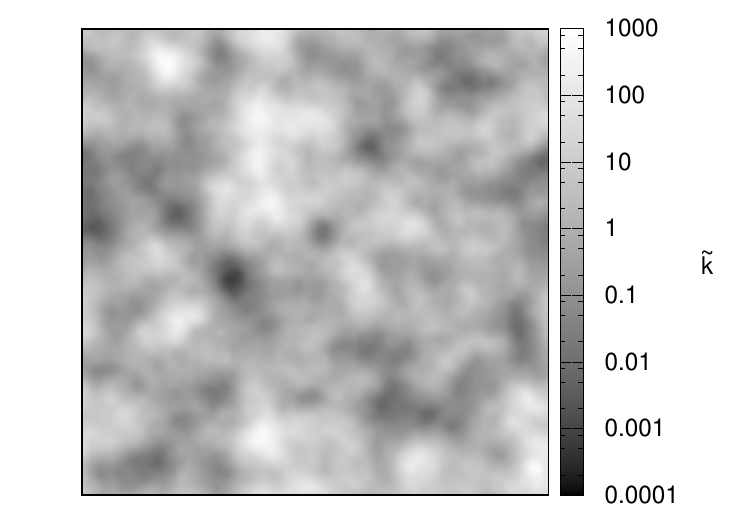}\label{Fig:perm}}
\subfloat[]{\includegraphics[width=.45\textwidth]{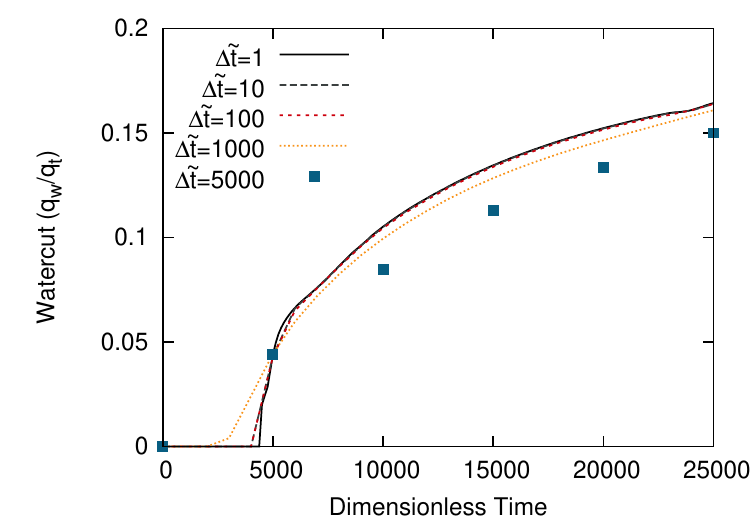}\label{Fig:watercut}}\\
\subfloat[$\tilde t=500$]{\includegraphics[width=.15\textwidth]{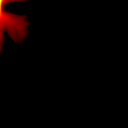}}
\hspace{.01\textwidth}
\subfloat[$\tilde t=5000$]{\includegraphics[width=.15\textwidth]{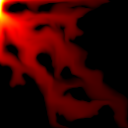}}
\hspace{.01\textwidth}
\subfloat[$\tilde t=10000$]{\includegraphics[width=.15\textwidth]{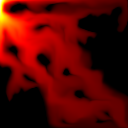}}
\hspace{.01\textwidth}
\subfloat[$\tilde t=15000$]{\includegraphics[width=.15\textwidth]{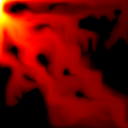}}
\hspace{.01\textwidth}
\subfloat[$\tilde t=25000$]{\includegraphics[width=.15\textwidth]{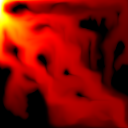}}
\hspace{.001\textwidth}
\includegraphics[height=.15\textwidth]{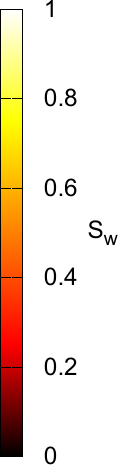}
 \caption{Simulation of two-phase flow in 2D using the CHD model as a homotopy map. The largest stable timestep for the traditional equations is $\Delta\tilde t=1$.  The homotopy approach allows the timestep to be increased by more than three orders of magnitude without significantly affecting the calculated watercut. (a) Plot of the permeability field, $\tilde k$. (b) Water saturation at the producing well as a function of time, calculated for several different $\Delta\tilde t$. (c)-(g) Snapshots of the simulation.}
 \label{Fig:2D-injection}
\end{figure*}

\begin{figure}[t]
\centering
\subfloat[intial]{\includegraphics[width=.15\columnwidth]{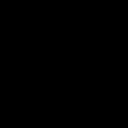}}
\hspace{.01\columnwidth}
\subfloat[1 iteration]{\includegraphics[width=.15\columnwidth]{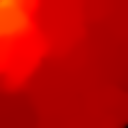}}
\hspace{.01\columnwidth}
\subfloat[2 iterations]{\includegraphics[width=.15\columnwidth]{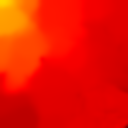}}
\hspace{.01\columnwidth}
\subfloat[3 iterations]{\includegraphics[width=.15\columnwidth]{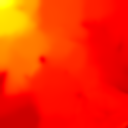}}
\hspace{.01\columnwidth}
\subfloat[4 iterations]{\includegraphics[width=.15\columnwidth]{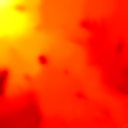}}
\hspace{.005\columnwidth}
\includegraphics[height=.15\columnwidth]{colorbar.pdf}\\
\subfloat[5 iterations]{\includegraphics[width=.15\columnwidth]{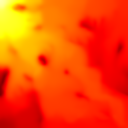}}
\hspace{.01\columnwidth}
\subfloat[6 iterations]{\includegraphics[width=.15\columnwidth]{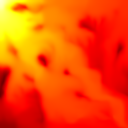}}
\hspace{.01\columnwidth}
\subfloat[7 iterations]{\includegraphics[width=.15\columnwidth]{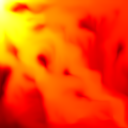}}
\hspace{.01\columnwidth}
\subfloat[8 iterations]{\includegraphics[width=.15\columnwidth]{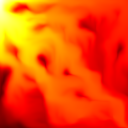}}
\hspace{.01\columnwidth}
\subfloat[9 iterations]{\includegraphics[width=.15\columnwidth]{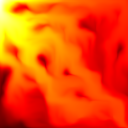}}
\hspace{.005\columnwidth}
\includegraphics[height=.15\columnwidth]{colorbar.pdf}
 \caption{Progression of the homotopy solution during one timestep with a very large $\Delta\tilde t=10^6$. Each frame shows the solution after successive multigrid iterations where $\gamma$ is decreased by a factor $w=.25$ after each iteration.}
 \label{Fig:homotopy}
\end{figure}

Porous flow through a highly heterogeneous permeability field in 2D provides a more realistic and rigorous comparison of the phase-field model and the classical model. A heterogeneous permeability field, shown in Fig. \ref{Fig:perm}, was generated using a modified exponential covariance function \cite{Gelhar1983} with an autocovariance of $s=5$ and a correlation length of 5 grid cells. Water was injected at the top left cell, and an oil/water mix was removed from the lower right cell. The results of the simulation are shown in Fig. \ref{Fig:2D-injection}c-g. Additional simulation parameters are listed in Table \ref{Tbl:parameters}, where tilde notation indicates dimensionless variables. The nondimensionalization scheme can be found in \ref{Sec:Nondimensionalization}.

This problem, although conceptually simple, was designed to place a severe timestep restriction on the Buckley-Leverett equations given by Eq. \ref{Eq:BL}. The maximum stable timestep is approximately $\Delta\tilde t=1$, and thousands of timesteps are required to simulate water breakthrough. Choosing larger time steps results in a solution that does not converge. In contrast, the CHD model discretized with a convex energy splitting (Eq. \ref{Eq:discretization}) converges for large $\Delta\tilde t$ when $\tilde\gamma$ is sufficiently large, and it was found that the timestep can be increased to $\Delta\tilde t=10^8$ without a loss of convergence. With such a large $\Delta\tilde t$ however, the reservoir goes from full to completely flooded in just one timestep, and important flow phenomena are completely over-stepped.

Figure \ref{Fig:2D-injection}(b) shows that, for practical purposes, the timestep may be increased by three to four orders of magnitude using the homotopy method without significantly affecting the calculated water cut at the producer, $\frac{q_w}{q_t}$, a main quantity of interest in reservoir simulation. Further increases in $\Delta\tilde t$, while stable, do not provide sufficient temporal resolution to capture water breakthrough. Truncation error introduced by the first-order time integration scheme may also be a concern, but could be addressed with an adaptive timestepping scheme. Removing the timestep restriction means that $\Delta\tilde t$ can be chosen as desired to balance the running time of the simulation and the timestep truncation error incurred.

Figure \ref{Fig:homotopy} shows the progression of the homotopy method during the first nine iterations of the first timestep of the 2D test problem with $\Delta\tilde t=10^6$. The progression illustrates how the homotopy method helps to find a guess that eventually converges to the solution. The initial iteration goes from full to flooded everywhere, and then as $\tilde\gamma$ is decreased over successive iterations the detailed structure of the flow emerge.

\begin{figure}[t]
\centering
\includegraphics[width=.5\columnwidth]{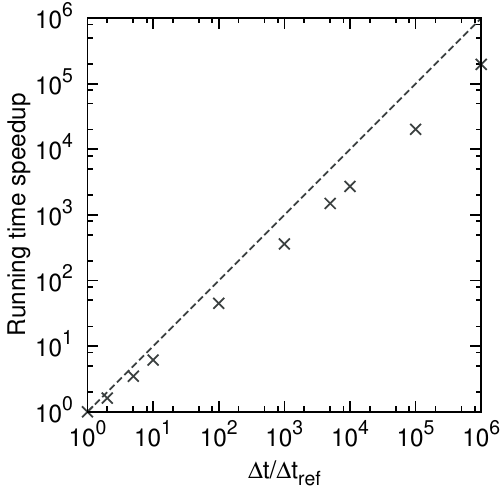}
 \caption{Computational speedup with increasing $\Delta t$ for the test problem in Figure \ref{Fig:2D-injection}. The dashed reference line indicates linear scaling of speedup with the timestep increase.}
 \label{Fig:speedup}
\end{figure}

Finally, the effect of homotopy on the computational running time of the simulation is of practical importance. Although the homotopy method permits much larger timeteps, more iterations are required for convergence. The running time of the solver was measured for different $\Delta\tilde t$, using relative error, $\|d\|/\|d_0\|<10^{-6}$, as a stopping criteron for the multigrid iterations, where $\|d_0\|$ is the $L_1$ norm of the initial defect.  Figure \ref{Fig:speedup} compares the computational speedup to the relative increase in $\Delta\tilde t$, and shows a nearly linear speedup with $\Delta t$ over six orders of magnitude.

\section{Conclusion}
Despite decades of sophisticated improvements to non-linear solvers, timestep restrictions during the simulation of porous flow remains a limiting factor in the simulation of large oil reservoirs. In this work it was shown that the restriction arises from limitations with Newton's method, which is not always guaranteed to converge, and that a viable solution is to convexify the flow equations, which obviates the need for specialized solvers. Two-phase incompressible flow without capillary potential was solved for large timesteps by introducing a macroscopic surface tension and employing a convex energy splitting scheme, following the phase-field approach. The numerical advantages of the modified model were investigated, and it was shown to be robust to high degrees of heterogeneity in permeability and viscosity contrast. Using a homotopy approach where the phase-field solution is continuously transformed into the underlying fractional flow solution by progressively decreasing the surface tension, the maximum stable timestep for a 2D simulation of fluid displacement could be increased by more than four orders of magnitude. We expect  that this homotopy scheme will be valuable for the petroleum migration stage of basin simulation, and because phase-field models can be derived for multiphase flow, anticipate that the method will be extendable to more complicated reservoir models such as three-phase flow, dual-permeability dual-porosity, and black oil.

\section*{Acknowledgments}
We would like to thank S.M. Wise for kindly answering questions about his unconditionally energy-stable discretization of the Cahn-Hilliard-Darcy system, and X. Liang for valuable discussion about the convergence of Newton's method. We thank A.H. Dogru of Saudi Aramco for his support and encouragement. This work was supported by Saudi Aramco's EXPEC Advanced Research Center.

\appendix
\section{Nondimensionalization}
\label{Sec:Nondimensionalization}
The evolution equations  are nondimensionlized by defining a porosity $\phi$, a time scale $\tau$, a length scale $L$, and a characteristic pressure $p_0$, which has units of energy density since one $Pa$ is a $J/m^3$. The nondimensional parameters are show in the following table:
\begin{center}
 \begin{tabular}{cccc}
  \hline
  Variable & Symbol & SI Units & Nondimensionalization \\
  \hline
  time & $t$ & $s$ & $\tilde t=\frac{t}{\phi\tau}$ \\
  pressure & $p$ & $Pa$ & $\tilde p=\frac{p}{p_0}$ \\
  permeability & $k$ & $m^2$ &  $\tilde k=\frac{k}{L^2}$ \\
  viscosity & $\eta$ & $Pa\cdot s$ &  $\tilde\eta=\frac{\eta}{p_0\tau}$ \\
  gradient energy & $\kappa$ & $J/m$ & $\tilde \kappa=\frac{\kappa}{p_0L^2}$ \\
  barrier height & $H$ & $J/m^3$ & $\tilde H=\frac{H}{p_0}$ \\
  source & $q_{ij}$ & $m^3/s$ & $\tilde q_{ij}=\frac{\phi\tau}{L^3}q_{ij}$ \\
  Dirac delta & $\delta(\vec x)$ & $m^{-3}$ & $\tilde\delta(\vec x)=L^3\delta(\vec x)$ \\
  \hline
 \end{tabular}
 \end{center}

\bibliographystyle{elsarticle-num}
\bibliography{phase-field-fluids}
\end{document}